\documentclass[aps,prl,twocolumn]{revtex4}
\usepackage{amsmath,bm,epsfig}
\usepackage{amssymb}
\newcommand{\B}[1]{{\bm{#1}}}
\newcommand{\C}[1]{{\mathcal{#1}}}    

\begin{document}
\title{Direct Estimate of the Static Length-Scale Accompanying the Glass Transition}
\author{Smarajit Karmakar, Edan Lerner, and Itamar Procaccia}
\affiliation{Dept. of Chemical Physics, The Weizmann Institute of
Sceince, Rehovot 76100, Israel}

\begin{abstract}
Characterizing the glass state remains elusive since its distinction from a liquid state is not obvious. Glasses are
liquids whose viscosity has increased so much that they cannot flow. Accordingly there have been many attempts to define a static length-scale associated with the dramatic slowing down of supercooled liquid with decreasing temperature. Here we present a
simple method to extract the desired length-scale which is highly accessible both for experiments and for numerical simulations. The fundamental idea is that low lying vibrational frequencies come in two types, those related to elastic response and those determined by plastic
instabilities. The minimal observed frequency is determined by one or the other, crossing at a typical length-scale which is growing with
the approach of the glass transition. This length-scale characterizes the correlated disorder in the system, where on longer length-scales the details of the disorder become irrelevant,  dominated by the Debye model of elastic modes.
\end{abstract}
\maketitle
\begin{figure*}
\includegraphics[scale = 0.40]{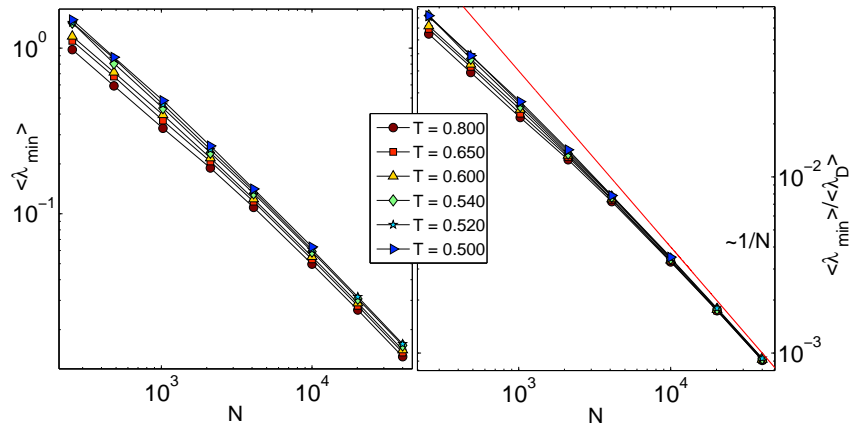}
\caption{Left Panel : Minimal eigenvalue calculated for a model glass former interacting via a pure
repulsive pair-wise potential plotted as a function of system size. Right Panel: Minimal eigenvalue
rescaled by the high frequency shear modulus ( which is proportional to the square of the Debye frequency) plotted
as a function of system size. Notice that at larger system size the rescaled eigenvalue becomes independent
of temperature indicating the crossover to the continuum elasticity limit.}
\label{lammin2D}
\end{figure*}
The phenomenon of the glass transition in which super-cooled liquids exhibit a dramatic slowdown in their dynamics upon
cooling is attracting a tremendous amount of effort. One of the thorny issues has to do with attempts to define and measure
a {\em static} length-scale that will grow upon the approach to the glass transition. Several such attempts were published
in the recent literature, using delicate measures of higher-order correlation functions\cite{05Chi4,09KDS}, of the effects
 of boundary conditions, \cite{08BBCGV}, point to set correlations \cite{07FM}, the scaling of the non affine displacement field \cite{10MGIO} and patch correlation scale \cite{09KL}. In this
 Letter we propose a different approach which yields a very natural static length that appears to fit the bill. An advantage
 of this approach is that it can be immediately and easily applied to both experimental \cite{10GCSKB} and numerical
 data (see below).

Our starting point is the fact that at lower frequency the density of state (DOS) for a disorder system
reflects the access of plastic modes which are not exhibited by the density of states of the corresponding crystalline materials \cite{02TWLB,10Sok}. This excess of modes is sometime referred to as the `Boson Peak' \cite{09IPRS}. It is most natural to discuss the `modes' in terms of the eigenfunctions of the Hessian
\begin{equation}
H_{ij}\equiv \frac{\partial^2 U(\B r_1,\B r_2, \dots \B r_N)}{\partial \B r_i\partial \B r_j} \ ,
 \end{equation}
where $U(\B r_1,\B r_2, \dots \B r_N)$ is the total energy of the system consisting of $N$ particles, and $\{\B r_k\}_{k=1}^N$ are
 their coordinates. Being real and symmetric, the Hessian is diagonalizable,  and in all discussions below we refer to this matrix with the Goldstone modes (zero modes due to symmetries) being pruned out. Recently we had progress in understanding the analytic form of the density of states of generic glassy systems \cite{11HKLP}. The eigenvalues appear in two distinct families, one corresponding to the Debye model
 of an elastic body, while the density of access plastic modes could be approximated as
 \begin{equation}
 f_{\rm pl}(\lambda) = B(T)\lambda^\theta \ ,  \text{for}~ \lambda\to 0 \ , \label{fpl}
 \end{equation}
 where the pre-factor $B(T)$ being strongly dependent on temperature and the exponent $\theta$ being weakly dependent on temperature. This
  dependence is a partial measure of the degree of disorder which grows with temperature, or for quenched systems, is partially controlled by the quench protocol from the melt to the glass. Together with the standard Debye contribution one can approximate the low-frequency tail of the density of states with the following approximation
\begin{equation}
P(\lambda) \simeq A\left( \frac{\lambda}{\lambda_D} \right)^{\frac{d-2}{2}}
+ B(T) f_{\rm pl}\left( \frac{\lambda}{\lambda_D} \right) \ . \label{Poflam}
\end{equation}
The first term , which is the Debye approximation of the DOS for a continuum elastic medium, contains
$\lambda_D \simeq \mu \rho^{2/d - 1}$, which is the Debye frequency and $\mu$ is the shear modulus.  For the sake of generality  we do not assume here any form for the distribution $ f_{\rm pl}$, but we will see bellow that the form (\ref{fpl}) will reappear naturally in the analysis below.

The main idea which will flash out the static typical scale is that the {\em minimal} eigenvalue $\lambda_{\rm min}$ observed in a system
of $N$ particles will be determined by the first {\em or} the second term in Eq. \ref{Poflam}. There will be a value of $N$
where a cross-over will occur, such that for larger system the Debye form determines $\lambda_{\rm min}$ whereas for small $N$
the plastic density of states determines $\lambda_{\rm min}$. This cross-over will be interpreted below in terms of a typical
length-scale separating correlated disorder from asymptotic elasticity.

To see the cross over integrate Eq. (\ref{Poflam}) from zero to $\langle \lambda_{\rm min}\rangle $. Since every quenched
system is random, every realization will have a somewhat different value of $\lambda_{\rm min}$, and we therefore consider
an ensemble average over many realization of system of the same number of particles $N$ quenched with the same protocol.

\begin{equation}
 N\int_{0}^{\left<\frac{\lambda_{\rm min}}{\lambda_{\rm D}}\right>} P\left(\frac{\lambda}{\lambda_D}\right)d\left(\frac{\lambda}{\lambda_D}\right) = 1.
\end{equation}
Introducing Eq. (\ref{Poflam}) into the integral we find
\begin{equation}
 \frac{Ad}{2} \left<\frac{\lambda_{\rm min}}{\lambda_{\rm D}}\right>^{d/2}
+ B(T)\int_{0}^{\left<\frac{\lambda_{\rm min}}{\lambda_{\rm D}}\right>} f_{pl}(\lambda)d\lambda =
\frac{1}{N} \ .
\end{equation}
Defining
\begin{equation}
{\cal G}\left(\left< \frac{\lambda_{\rm min}}{\lambda_{\rm D}} \right>\right) \equiv \int_{0}^{\left<\frac{\lambda_{\rm min}}{\lambda_{\rm D}}\right>} f_{pl}(\lambda)d\lambda \ ,
\end{equation}
we end up with
\begin{equation}
{\cal G}\left(\left< \frac{\lambda_{\rm min}}{\lambda_{\rm D}} \right>\right) = \left[ \frac{1}{B(T)}\left(\frac{1}{N} -
\frac{Ad}{2} \left<\frac{\lambda_{\rm min}}{\lambda_{\rm D}}\right>^{d/2} \right) \right] \ .
\end{equation}
Changing slightly this equation, writing  $N=\rho V$, and renormalizing the coefficient $A$ accordingly,
we write
\begin{equation}
{\cal G}\left(\left< \frac{\lambda_{\rm min}}{\lambda_{\rm D}} \right>\right) = \left[ \frac{1}{\rho B(T)}\left(\frac{1}{V} -
\frac{\tilde A d}{2} \left<\frac{\lambda_{\rm min}}{\lambda_{\rm D}}\right>^{d/2} \right) \right] \ .
\end{equation}
Finally, since the function $\C G$ is monotonically increasing,  $\left<\frac{\lambda_{\rm min}}{\lambda_{\rm D}}\right>$ can be written as a scaling function of the form
\begin{equation}
\left< \frac{\lambda_{\rm min}}{\lambda_{\rm D}} \right>={\C F}\left[ \xi^d(T)\left(\frac{1}{V} -
\frac{\tilde A d}{2} \left<\frac{\lambda_{\rm min}}{\lambda_{\rm D}}\right>^{d/2} \right) \right].
\label{ansatz}
\end{equation}
where $\C F\equiv \C G^{-1}$ and $\xi^d(T) \equiv \frac{1}{\rho B(T)}$. The typical scale $\xi(T)$ will be calculated by demanding that all the data calculated for
different system sizes and temperatures should
collapse into a master curve just by appropriately choosing the $\xi(T)$.

The analysis is then done as follows. Two typical glass formers were considered, both obtained from a binary mixture, one
with a purely repulsive potential and the other with repulsive and attractive contribution (the Kob-Andersen model). The details of the potentials can be found in \cite{10KLP,95KA}. The systems were equilibrated at some temperature $T> T_g$ and then instantly quenched by direct
energy minimization to the nearest local minimum of the energy landscape. At this state the Hessian was computed and the minimal
eigenvalue was obtained using the Lanczos algorithm \cite{wik}. For a given system size $N$ and temperature $T$ this procedure was repeated to have an average $\langle \lambda_{\rm min}\rangle$ until convergence was achieved. At this point the temperature or the system
size were changed and the procedure was repeated, to eventually have a table of $\langle \lambda_{\rm min}\rangle(N,T)$. The density
was fixed for every type of potential, (0.85 for pure repulsion in 2$D$, 0.82 for pure repulsion in 3$D$, and 1.20 for the Kob-Andersen model).

In the left Panel of Fig~\ref{lammin2D}, we have plotted the minimal eigenvalue $\left<\lambda_{\rm min}\right>$ of the
two-dimensional purely repulsive super-cooled system as a
function of system size $N$ for different temperatures. In the right Panel the same minimal eigenvalue
was rescaled by the characteristic Debye value $\lambda_{\rm D}$ and plotted as a function of system
size. Note that at larger system size they become independent of temperature consistent with the fact that
at larger system size one recovers trivial scaling predicted by the continuum elasticity theory.

\begin{figure}
\includegraphics[scale = 0.33]{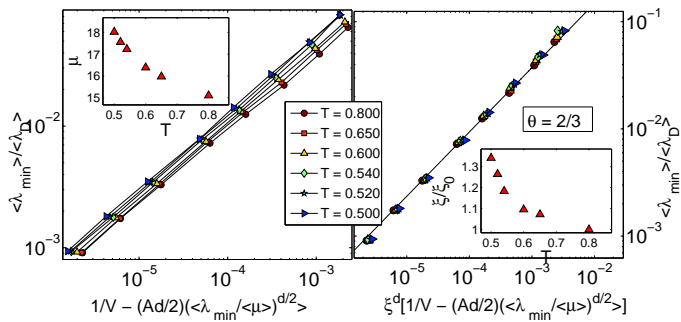}
\caption{Scaling of the minimal eigenvalue calculated for a model glass former interacting via a pure
repulsive pair-wise potential in two dimensions. here and in all the figures below $\xi$ was chosen as $\xi=\xi_0=1$ at
the highest available temperature. }
\label{lammin2DScaling}
\end{figure}

\begin{figure}
\includegraphics[scale = 0.35]{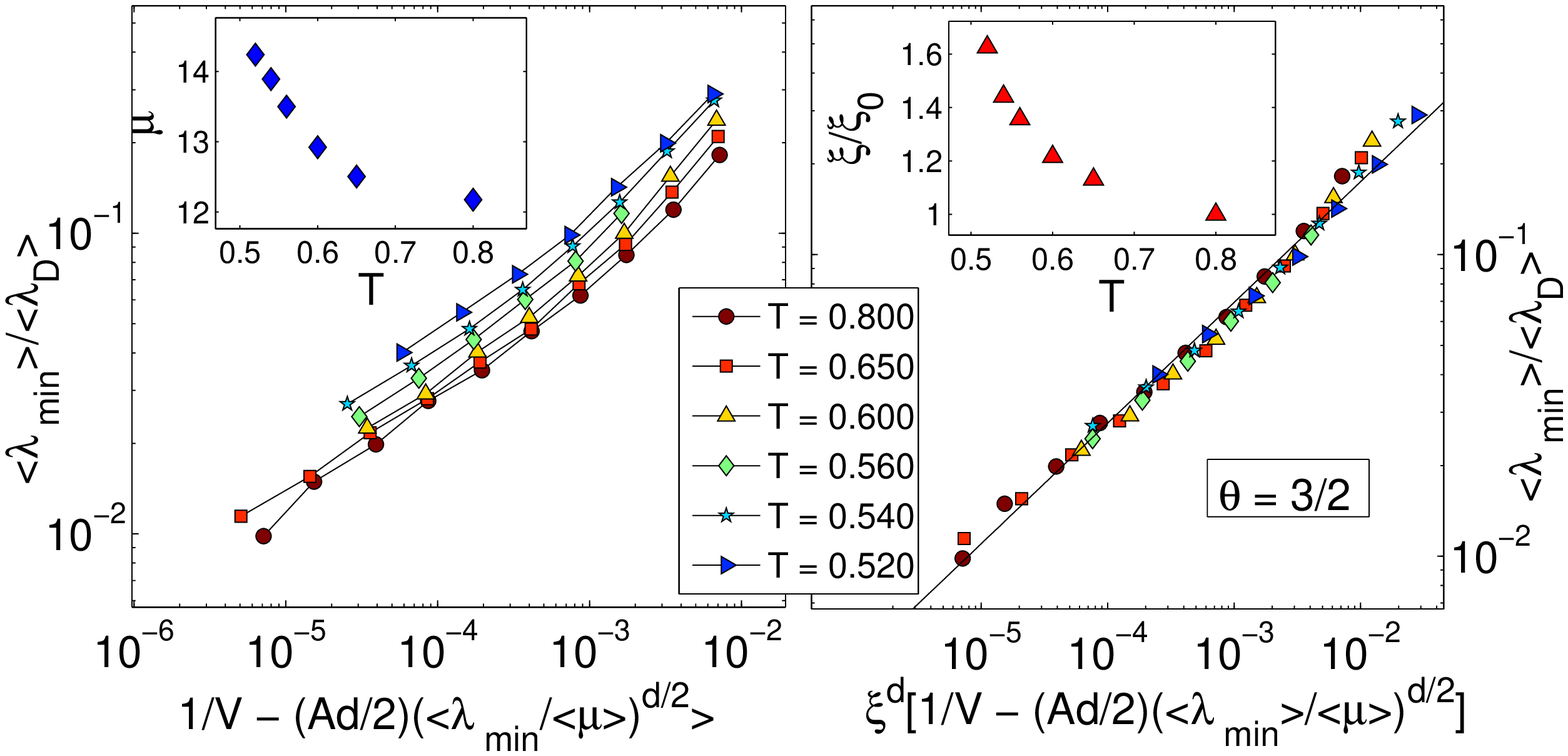}
\caption{The measured minimal eigenvalue as a function of system size for a binary system interacting via a pure
repulsive potential in 3 dimension. Every data point represents an average over 2000 Inherent structures. }
\label{lammin3D}
\end{figure}

\begin{figure}
\includegraphics[scale = 0.30]{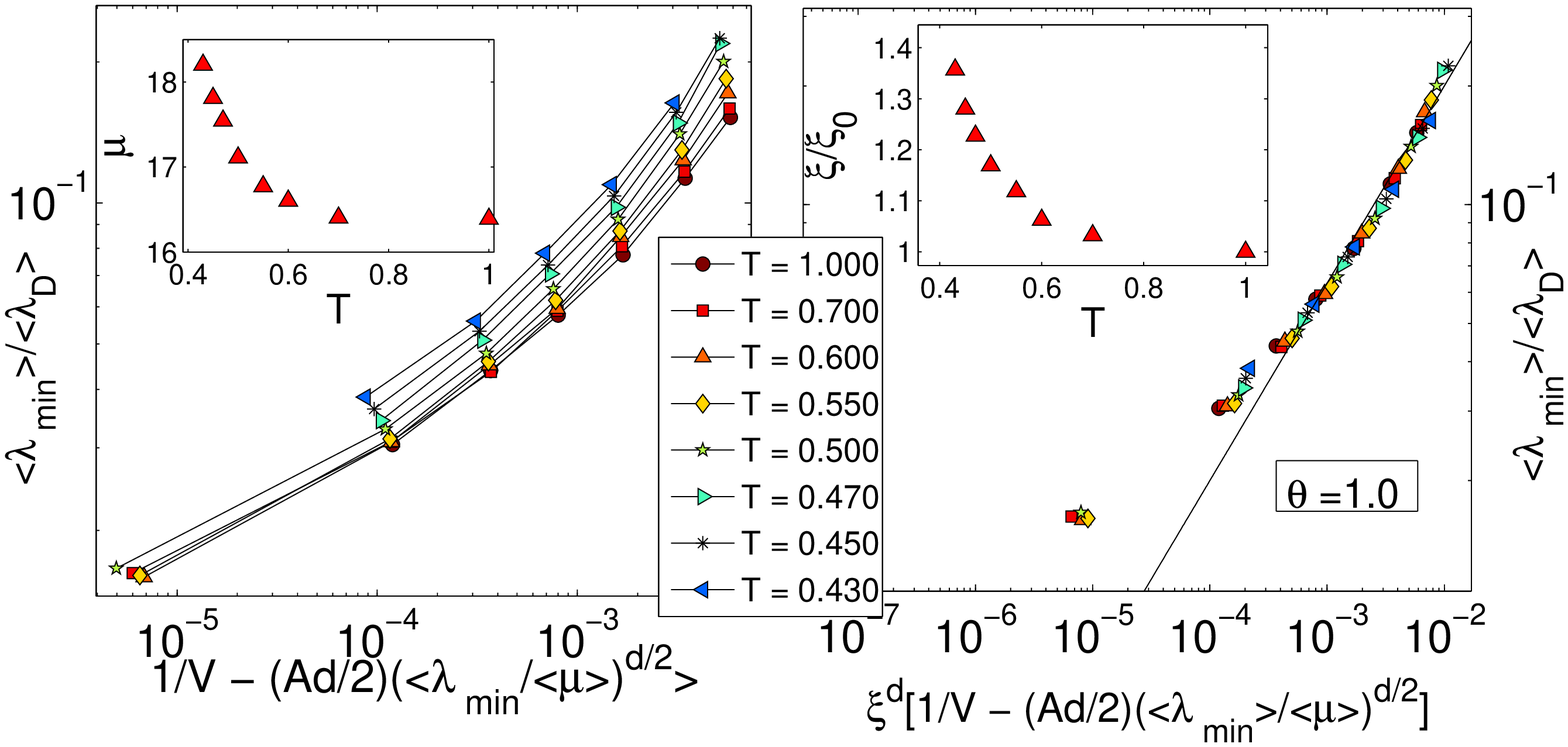}
\caption{The measured minimal eigenvalue as a function of system size for a binary system interacting via
Kob-Andersen potential in 3 dimension. Every data point represents an average over 2000 Inherent structures.}
\label{lamminKA}
\end{figure}

In Fig.~\ref{lammin2DScaling}, we re-plotted the same data of Fig. \ref{lammin2D} according to the scaling ansatz Eq.~\ref{ansatz}. The left panel exhibits the results without rescaling by $\xi^2(T)$. In the right panel we extracted the disorder length scale $\xi(T)$ by collapsing the data. The resulting data collapse into a straight line indicates that Eq. (\ref{fpl}) is obeyed to high precision. The collapse itself supports
the scaling ansatz Eq.~\ref{ansatz}. In the inset at the right panel we show how the typical scale increases when the glass transition is
approached.
In Fig, \ref{lammin3D} and \ref{lamminKA} we show the same the same analysis for the purely repulsive glass in 3D, a
and for the Kob-Andersen model in 3D. Both the data collapse and the resulting increase in the typical scale appear
very encouraging and support the approach proposed above.
\begin{figure}
\includegraphics[scale = 0.30]{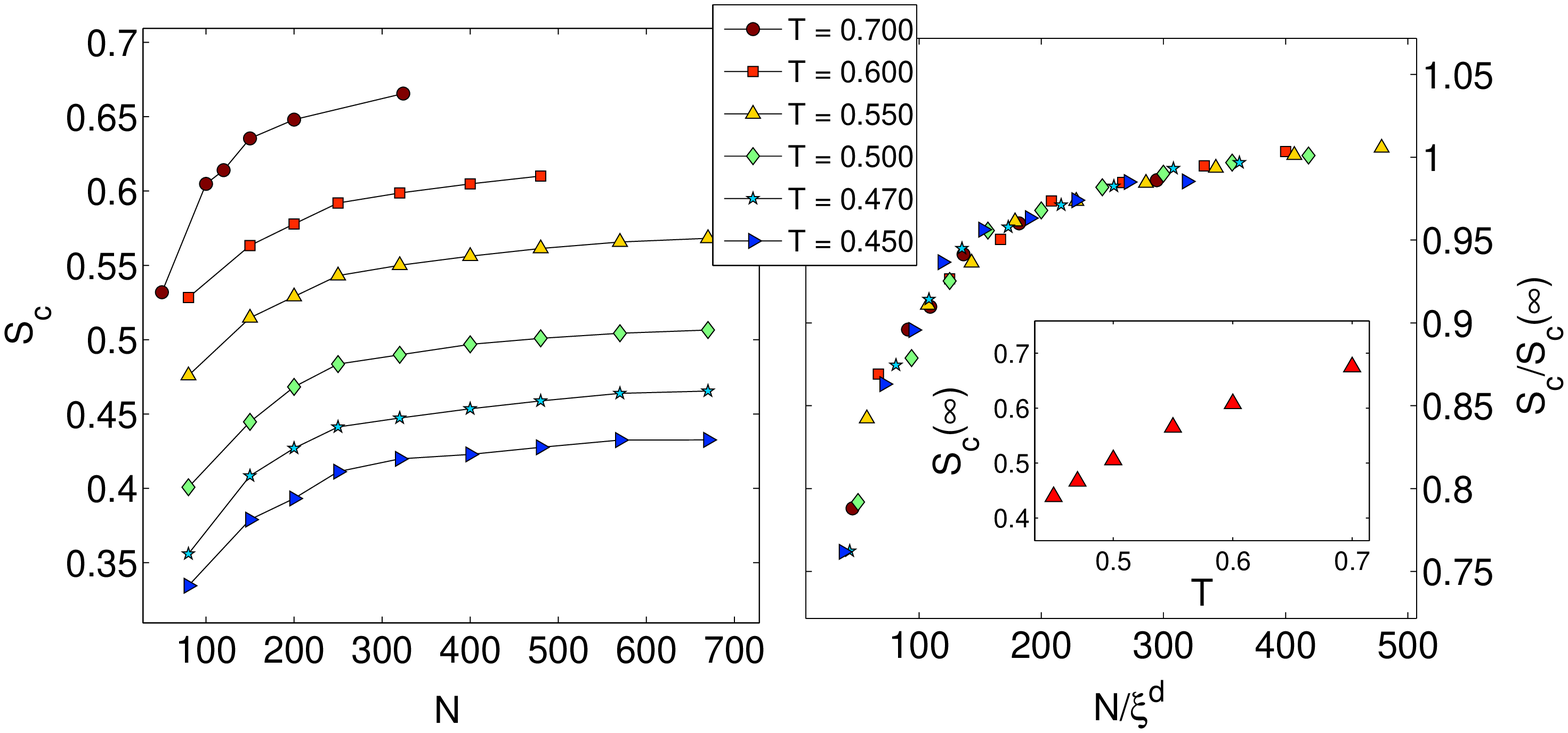}
\caption{ The scaling of configurational entropy with the length scale extracted from the finite size scaling of
minimal eigenvalue for the Kob-Anderson Model. The data is taken from \cite{09KDS,SmarajitThesis2009}}.
\label{entropy}
\end{figure}
Note that the approximate linearity in the collapsed scaling function in the right panel of Fig. \ref{lammin3D} again
indicates the relative accuracy of the functional form (\ref{fpl}) for the density of plastic modes. We have written the value of appropriate exponent $\theta$ in Eq. (\ref{fpl}) directly into the graphics. In the right panel
of Fig. \ref{lamminKA} we see some curvature that indicates that higher order terms in the density of states already
play some role. Nevertheless the data collapse is superb.

Finally, it is interesting to ask how our typical scale helps in understanding other measures of disorder that
were proposed in the past. As an example we consider here the configurational entropy $S_c(T)$ of the Kob-Andersen model
in 3D. This entropy was computed in Refs. \cite{09KDS,SmarajitThesis2009} and the reader is referred to these
publications for a full description of the method and the results. The configurational entropy is expected to
characterize the degree of disorder in the supercooled liquid. Once $S_c(T)$ was computed, it was discovered
phenomenologically that the data at different temperatures could be collapsed by rescaling the system size by
a typical scale $\ell (T)$. It remains however mysterious what that length-scale might be, and how to measure
it independently of the configurational entropy. At this point we can offer a resolution of that mystery.
In Fig. \ref{entropy} we show, in the left panel, the configurational entropy measured at different temperatures
as a function of the system size.  Since the number of accessible minima reduces as a function of temperature,
(high energy 'states' are excluded when the temperature decreases), the configurational entropy goes down as
seen in the left panel of Fig. \ref{entropy}. In the right panel we exhibit $S_C(T)/S_C(N\to \infty)$ as a function
of the re-scaled system size $N/\xi^3(T)$. The rescaling is achieved using our data for $\xi(T)$ as shown in the
right panel of Fig. \ref{lamminKA} without any re-fitting. The data collapse shows that the mysterious $\ell(T)$ is
precisely our typical scale as propose in this Letter.

We should stress that the estimate of our typical scale is easily achieved also in appropriate experiments. It
was shown in Ref. \cite{10GCSKB} that the density of states can be calculated in experiments on colloidal glassy
systems, and in particular $\lambda_{\rm min}$ is available. We propose that such measurements should be repeated as
a function of system size at different values of the packing fraction. Such data can be used as explained above to
determined the dependence of our typical scale on the packing fraction, throwing new light on the interesting
physics of these complex systems.
\begin{figure}
\includegraphics[scale = 0.40]{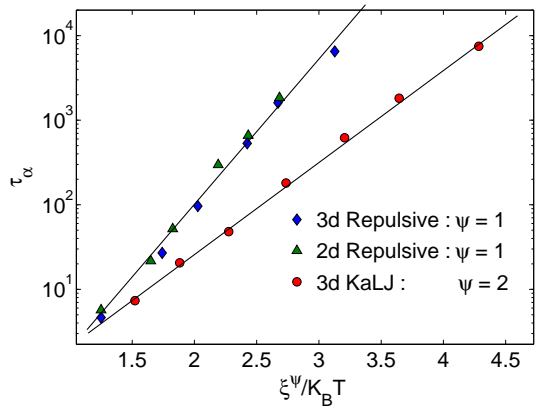}
\caption{The logarithm of the relaxation time $\tau_\alpha(T)$ measured for all the three models discussed above, plotted
against $\xi^\psi/T$. The straight lines are a guide to the eye.}
\label{txi}
\end{figure}

Finally, it is impossible to end this Letter without asking how the newly found length-scale correlates with the
dramatic slowing down that is observed in our super-cooled liquids as the temperature decreases. If we take the point
of view that the typical size of relaxation events also depends on $\xi$, either proportional to $\xi$ if they
are stringy in nature or on $\xi^2$ if they are planar, then we expect the relaxation time $\tau_\alpha$ to be of the order of
\begin{equation}
\tau_\alpha \propto \exp {[C \xi^\psi/k_B T}] \ , \label{tauvsxi}
\end{equation}
where $C$ is some unknown scale and $\psi$ is 1 or 2 depending on the geometry of the relaxation events. We see
that Eq. (\ref{tauvsxi}) is validated with $\psi=1$ for the repulsive model in both 2D and 3D, and with $\psi=2$ for the
Kob-Andersen model. It is not known at this point why indeed $\psi$ may differ when the attractive part of the potential
is added, and we must leave this issue for further research in the future, and see Ref. \cite{09BT,10PSD} for some comments on this issue.

In summary, we have proposed here a very simple method to extract the typical scale that separates a disorder dominated
regime from an elastic dominated behavior. All that is needed is the measurement of the minimal eigenvalue of the Hessian matrix
(or, equivalently, the minimal harmonic frequency of the system), for systems of different size and temperature. A simple
re-plotting procedure of the data is then used to extract the typical scale. This scale appears to properly collapse
the data of the configurational entropy, resolving a riddle that existed for some time regarding the nature of the
length-scale that does it. Finally, and only in a manner of passing, we also considered the relation of the obtained
length-scale to the relaxation time of the super-cooled liquids, and found a strong indication that this
length scale also determines the observed dynamics. We trust that the newly proposed length-scale would be computed
in further numerical and laboratory experiments by other groups to enhance the understanding of the glass transition.
In particular it would be useful to compare this length-scale to other length-scales that were proposed by other groups
as mentioned in the introduction.

This work had been supported in part by an ERC ``ideas" grant, the Israel Science Foundation and by the German Israeli Foundation.
A discussion with Satya Majumdar is acknowledged.

\end{document}